%% file: main.tex
\documentclass[preprint,12pt]{elsarticle}
\usepackage[T1]{fontenc}
\usepackage{amsmath,amssymb,booktabs,array,multirow,graphicx,xcolor,hyperref,tabularx}


\usepackage[capitalise]{cleveref}
\graphicspath{{figures/}}
\newcommand{\sys}{NGAC-IR}
\newdefinition{definition}{Definition}

\makeatletter\def\ps@pprintTitle{\let\@oddhead\@empty\let\@evenhead\@empty\def\@oddfoot{\footnotesize\itshape Preprint \hfill \today}\let\@evenfoot\@oddfoot}\makeatother

\begin{document}
\begin{frontmatter}
\title{Extracting Next Generation Access Control Policies from Natural Language Using LLMs: An Intermediate-Representation Approach with Back-Translation Verification}

\author[npu]{Ayesha Habib}
\author[npu]{Huixiang Zhang\corref{cor}}
\author[npucs]{Waiz Haider}
\cortext[cor]{Corresponding author}
\affiliation[npu]{organization={School of Cybersecurity, Northwestern Polytechnical University},
  city={Xi'an}, postcode={710129}, country={China}}
\affiliation[npucs]{organization={School of Computer Science, Northwestern Polytechnical University},
  city={Xi'an}, postcode={710129}, country={China}}

\begin{abstract}
Organizations write access-control rules in natural language, but NIST's Next Generation Access Control (NGAC) needs them as a formal policy graph. Large language models (LLMs) can automate this translation, but their output can break the format, add rules the text does not state, or drop conditions. We present \sys{}, which follows model-driven engineering (MDE). First, the LLM fills in an intermediate representation (IR), a JSON template that records who may, must or must not do what, on which resource and under which condition; a grammar keeps the output within this template. Second, a compiler that uses no LLM converts the IR into an NGAC policy graph and links each rule to its source sentence. Third, a back-translation check writes each rule in English and asks an LLM whether the source sentence states it. On four public datasets, all outputs could be compiled, compared with 95.5\% without the grammar, and 95.7\% were faithful to their sentence. \sys{} outperforms two published methods on all four datasets: the LLM-based Code4Policy in exact match and faithfulness, and the rule-based extractor of Abdelgawad et al.\ in rule type. The back-translation check sent 10.5\% of sentences to human review and found 26 of the 31 errors. To test enforcement, we deployed the compiled policies in NIST's Policy Machine: they gave the expected decision for all 390 access requests, and the tested obligations were carried out. On one local server with open-weight models, all 714 sentences take about 20 minutes and 0.35\,kWh, at zero API cost.
\end{abstract}

\end{frontmatter}

\section{Introduction}\label{sec:intro}

Hospitals, universities and companies write down who may do what in plain text: requirement documents, use cases and policy manuals. To check every access automatically, a computer system needs the same rules in a formal form. Next Generation Access Control (NGAC), a standard of the US National Institute of Standards and Technology (NIST), stores a policy as a graph: users, resources and their attributes are the nodes, and the edges are assignments (what belongs to what), associations (who may do what), prohibitions (who may not) and obligations (what the system must do when an event happens) \citep{nist800178}. Turning the text into such a graph by hand is slow and needs experts, and a single error can give access that nobody intended.

Large language models (LLMs) can now do this translation automatically \citep{lawal2024translating,nist2024translating,yang2025extraction,habib2026text2ngac} (\cref{sec:background}). Earlier methods are either not accurate enough on real documents or costly to run. The rule-based pipeline of \citet{abdelgawad2023synthesizing} uses no LLM and covers only associations, prohibitions and part of the assignments; on the four datasets used here, it finds the right rule type for only 33--74\% of sentences (\cref{sec:rq2}). Code4Policy \citep{lawal2024translating} uses an LLM, but it was tested on text written by an LLM, has no assignments and covers obligations only partly. Our earlier system Text2NGAC \citep{habib2026text2ngac} covers all four relation types and is accurate, but it relies on an ensemble of commercial LLMs (DeepSeek, GPT-4o-mini, Claude Sonnet and Claude Opus) reached through paid APIs. This costs about \$11--47 per 1{,}000 sentences, and every policy sentence is sent to an outside service. In this paper, we aim for the same quality with open-weight models on one local server, at zero API cost. All of these approaches also share two weaknesses. First, the LLM writes the rules as free text, so the output can break the required format. The flat rule format also cannot record a condition (``after registration closes''), how strong a rule is (``must'' or ``should''), or which sentence a rule came from. Second, the rules are usually checked only for their form, if they are checked at all. Text2NGAC, for example, runs five tests on the finished graph: every sentence must appear in it, every edge must link the right kinds of nodes, no rule may both allow and deny the same access, every obligation needs an event and a response, and no rule may appear twice. All five look at how the graph is built, not at what the rules say. Suppose a sentence lets a user see only the patient's medical identifier, but the rule gives access to the patient's whole record. The graph still passes all five tests. To measure this, Text2NGAC put such errors into correct graphs on purpose and counted how many the five tests caught: only 9.2\% \citep{habib2026text2ngac}.

There is also a problem with how such systems are scored. Usually, a system's rules are compared with the rules one annotator wrote, and only an exact match counts as correct. But one sentence can often be written correctly in more than one way: an annotator may name a resource \texttt{appointment\_detail} where a system writes \texttt{appointment\_details}, or record ``the system displays the list'' as an assignment where another reading gives an obligation. Exact match counts each such difference as an error, and even two careful annotators rarely write the same rules word for word: in an independent second coding of the datasets used here, two annotators did not even agree on the rule type of every sentence (Cohen's $\kappa = 0.84$) \citep{habib2026text2ngac}. We therefore use faithfulness as the main measure: whether the extracted rules say what their sentence says, checked against the sentence and not against the annotator's rules. The difference is large: the same outputs that match the annotation exactly for 30--64\% of sentences are faithful to their sentence for 95.7\% of all 714 sentences.

This paper addresses these problems with a model-driven engineering (MDE) approach \citep{schmidt2006mde,brambilla2017mde}, in which a formal model is filled in first and then converted with fixed rules. The LLM no longer writes the policy itself; it only fills in a small, fixed template (the model), and every later step either follows fixed rules or is checked separately. We call the resulting system \sys{}. We make four contributions.

\begin{enumerate}
\item \textbf{An IR approach: the LLM fills in a template, and a compiler builds the policy.} The LLM does not write the policy. For each sentence, it fills in a fixed template, the intermediate representation (IR): who may, must or must not do what, on which resource, and under which condition. A grammar keeps the LLM's output inside this template \citep{willard2023outlines,geng2023grammar}, so every output can be compiled. A compiler with fixed rules, which uses no LLM, then turns the IR into an NGAC policy graph and into Policy Machine Language. Every rule keeps the identifier of the sentence it came from.
\item \textbf{A back-translation check for meaning.} Each extracted rule is written back as an English sentence, and a separate LLM call checks whether the source sentence states this rule. The check sees the rule, the source sentence and the two sentences before it, but not the ground truth, so it can find wrong meanings that structural tests miss. Five simple rule checks, which use no LLM, cover cases the LLM step cannot see, such as a duty recorded only as a containment. A flagged rule is sent to a person for review, together with its sentence.
\item \textbf{A fairer way to measure quality.} Exact match counts an output as correct only if it equals one annotator's rules word for word. We instead check whether the extracted rules say what their sentence says (faithfulness). To check the faithfulness of all methods fairly, we marked their outputs for all 714 sentences blind: in shuffled order and without the name of the method that produced each output. We compare \sys{} with two published methods, our earlier system Text2NGAC and a rebuilt direct extractor, and show why exact match understates quality.
\item \textbf{Good results at zero API cost.} Text2NGAC needs paid commercial LLMs, and every policy sentence leaves the organization. \sys{} uses only open-weight models on one local server with two consumer GPUs, so it has no API cost and the policy text stays inside the organization. Our goal is to reach the quality of Text2NGAC in this setting.
\end{enumerate}

In short, \sys{}'s output is faithful to its sentence for 95.7\% of sentences, whereas the other methods reach 83.9--90.3\%. \sys{} makes errors on 31 sentences, less than half as many as our earlier system Text2NGAC (70). The grammar makes every output valid (without it, about one output in twenty is broken). The back-translation check finds errors of meaning that structural tests miss, and it wrongly flags only about one correct association in forty; on \sys{}'s own output it sends 10.5\% of sentences to review and finds 26 of the 31 errors, and every flag points to the sentence to reread. \sys{} is also more faithful than Text2NGAC (95.7\% compared with 90.2\%) without any commercial model.

The rest of the paper is organized as follows. \Cref{sec:background} gives background and related work. \Cref{sec:approach} presents the IR, the compiler and the checks. \Cref{sec:setup} describes the evaluation, \cref{sec:results} reports the results, and \cref{sec:discussion} discusses limitations and threats to validity.

\section{Background and related work}\label{sec:background}

\subsection{Background}

\paragraph{NGAC} We use the following definitions, based on NIST SP 800-178 \citep{nist800178,ferraiolo2015pm}. NIST's Policy Machine is the reference implementation of NGAC.

\begin{definition}[Policy elements]\label{def:elements}
An NGAC policy is defined over finite sets of users $U$, user attributes $\mathit{UA}$, resources $R$, resource attributes $\mathit{RA}$ and policy classes $\mathit{PC}$, and a set of operations $\mathit{Ops}$.
\end{definition}
\noindent User attributes group users, for example by role; resource attributes group resources, for example by type; policy classes are the top-level containers. NIST SP 800-178 calls resources \emph{objects} (O) and resource attributes \emph{object attributes} (OA).

\begin{definition}[Assignment]\label{def:assign}
The assignment relation is
\[
\begin{aligned}
\mathit{ASSIGN} \subseteq{} & (U \times \mathit{UA}) \cup (\mathit{UA} \times \mathit{UA}) \cup (\mathit{UA} \times \mathit{PC})\\
\cup{} & (R \times \mathit{RA}) \cup (\mathit{RA} \times \mathit{RA}) \cup (\mathit{RA} \times \mathit{PC}).
\end{aligned}
\]
A pair $(x, y) \in \mathit{ASSIGN}$ places $x$ inside $y$; the relation must be acyclic.
\end{definition}
\noindent The relation thus allows six kinds of assignment: U$\to$UA, UA$\to$UA, UA$\to$PC, R$\to$RA, RA$\to$RA and RA$\to$PC. For example, the user attribute doctor is assigned to medical staff, and the resource attribute patient record to medical data.

\begin{definition}[Association]\label{def:assoc}
An association $(\mathit{ua}, \mathit{ops}, \mathit{ra})$ gives the user attribute $\mathit{ua}$ permission to perform the operations $\mathit{ops}$ on the resource attribute $\mathit{ra}$. The permission holds for every user assigned to $\mathit{ua}$ and every resource assigned to $\mathit{ra}$, directly or through other attributes. Formally, $\mathit{ASSOC} \subseteq \mathit{UA} \times 2^{\mathit{Ops}} \times \mathit{RA}$, where $2^{\mathit{Ops}}$ is the set of all subsets of $\mathit{Ops}$.
\end{definition}
\noindent For example, the association (doctor, \{read\}, patient record) means that every doctor may read every patient record.

\begin{definition}[Prohibition]\label{def:prohib}
A prohibition $(\mathit{ua}, \mathit{ops}, \mathit{ra})$ has the same form as an association but denies the operations instead of granting them, for every user and resource below $\mathit{ua}$ and $\mathit{ra}$. A prohibition overrides any association that grants the same operations. Formally, $\mathit{PROHIB} \subseteq \mathit{UA} \times 2^{\mathit{Ops}} \times \mathit{RA}$.
\end{definition}

\begin{definition}[Obligation]\label{def:oblig}
An obligation is a pair $(\mathit{event}, \mathit{response})$: when an event that matches $\mathit{event}$ occurs, the system carries out $\mathit{response}$.
\end{definition}
\noindent For example, once a student has registered, the registration system sends information to the billing system.

\paragraph{Constrained decoding} Grammar-constrained or schema-constrained decoding restricts each generated token to those that keep the output inside a formal language \citep{willard2023outlines,geng2023grammar}. Serving engines such as vLLM \citep{kwon2023vllm} expose it through JSON Schema \citep{vllm2026structured}. It makes the output valid in form; whether it also makes it more accurate depends on the task, and we measure it.

\paragraph{Model-driven engineering} Model-driven engineering (MDE) builds software from models \citep{schmidt2006mde,brambilla2017mde}. A \emph{model} is a structured description, such as one IR document. Its \emph{metamodel} says what such a description may contain, such as the IR grammar; a model that follows its metamodel \emph{conforms} to it. A \emph{model transformation} turns one model into another with fixed rules \citep{czarnecki2006transformation}; our compiler is an example.

\paragraph{LLM judges} Using one LLM to grade the output of another is common, but the grades are not always right \citep{zheng2023judging}. We use an LLM judge only inside the back-translation check, where it compares one rule with one sentence. The quality of the extracted rules is measured by marking each output against its sentence (\cref{sec:setup}).

\subsection{Related work}

\paragraph{Early NLP methods} The first methods used classical natural language processing. Text2Policy \citep{xiao2012automated} finds the sentences in software documents that state access-control rules and extracts the rules from them (86.3\% accuracy), as well as the action steps of use cases. ACRE \citep{slankas2014relation} first finds the sentences that state access-control rules and then extracts the rules with patterns over the dependency parse of each sentence. \citet{narouei2018identification} use semantic role labeling to extract the parts of each rule from unrestricted natural-language documents (75\% F1), and \citet{alohaly2019deep} use deep learning to extract the attributes of subjects and objects for ABAC. Two of the datasets we use come from this line of work: the iTrust sentences of Text2Policy (T2P) and of ACRE. These methods produce flat subject--action--resource rules or single rule elements. They do not target NGAC: they have no hierarchy of users and resources and no obligations.

\paragraph{Abdelgawad et al.} \citet{abdelgawad2023synthesizing} were the first to extract NGAC policies from real policy documents. Their rule-based pipeline uses the spaCy dependency parser and a list of negation words to find associations, prohibitions and part of the assignments (users to user attributes and resources to resource attributes), and checks the resulting graph with database queries. They report about 75\% accuracy on seven university policy documents. The pipeline uses no LLM, handles no obligations, and turns conditions into plain associations. On the four datasets used here, it finds the right rule type for only 33--74\% of sentences (\cref{sec:rq2}).

\paragraph{Code4Policy} \citet{lawal2024translating} prompt GPT-4 to write NGAC relations as Python data classes, with and without examples, and check the entity types of the output; a NIST report by the same authors describes a related chain-of-thought approach \citep{nist2024translating}. They report 99\% F1, but on policy text that was itself written by an LLM, which lacks the compound sentences, chains of pronouns and ambiguous phrases of real documents. Code4Policy has no assignments, records obligations only partly as conditions, and does not check what the rules mean. On the four datasets used here, its exact match is 3.7--18.7\% (\cref{sec:rq2}).

\paragraph{Knowledge distillation for ABAC} \citet{yang2025extraction} train a small model on the output of a large LLM to extract ABAC rules that a system can enforce. The rules are flat subject--action--resource triples, so they have no NGAC hierarchy, prohibitions or obligations.

\paragraph{Text2NGAC} Our earlier system \citep{habib2026text2ngac} is, to our knowledge, the first to extract all four NGAC relation types from real documents. It first rewrites unclear sentences more clearly. One commercial LLM (DeepSeek) then decides the rule type of each sentence, and the sentences it is unsure about go to an ensemble of several commercial LLMs that vote, and Claude Opus decides the remaining cases. The rules are then extracted. Five structural tests check the graph, and mutation testing measures how many errors these tests catch. Text2NGAC is accurate, but it costs about \$11--47 per 1{,}000 sentences in API fees, every policy sentence is sent to an outside service, the LLM writes the rules as free text, and the structural tests miss errors of meaning. When a rule gives access to more than its sentence allows, for example to the patient's whole record when the sentence allows only the patient's medical identifier, the graph is still well formed; the tests caught only 9.2\% of such errors.

None of these systems constrains the model's output to a formal grammar, and none checks the meaning of an extracted rule against its source sentence. \sys{} does both, on open-weight models.

\section{Approach}\label{sec:approach}

\begin{figure}[t]
\centering
\includegraphics[width=\textwidth]{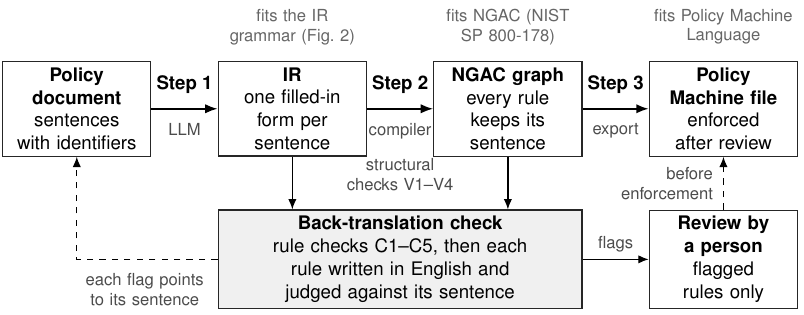}
\caption{The \sys{} pipeline. Steps 1--3 are the model transformations $T_1$--$T_3$; the gray notes name the metamodel each model must fit. The back-translation check reads the IR and the graph, and each flag points back to its source sentence.}
\label{fig:pipeline}
\end{figure}

\sys{} lets the LLM do only what it is good at, reading a sentence, and leaves everything else to code that follows fixed rules. \Cref{fig:pipeline} shows the pipeline. In Step~1, the LLM fills in a fixed form, the intermediate representation (IR), for each sentence. In Step~2, a compiler that uses no LLM turns these forms into an NGAC graph, in which every rule keeps the identifier of the sentence it came from, and then runs four structural checks, V1--V4, to make sure the graph is well formed. In Step~3, the graph is exported to a Policy Machine file, which at this point is only a draft. At the same time, the back-translation check looks at meaning: five rule checks, C1--C5, look for signs that the LLM misread a sentence, and an LLM step writes each rule in English and compares it with its sentence. A person reviews the flagged rules and either accepts the draft or corrects it, and only then is the policy enforced. In MDE terms, the IR documents, the NGAC graph and the Policy Machine file are models; the gray notes in the figure name the metamodel each one must fit, and the three steps are the model transformations $T_1$, $T_2$ and $T_3$.

\subsection{Step 1: the LLM fills in the IR}\label{sec:metamodel}\label{sec:t1}

\begin{figure}[!t]
\centering
\includegraphics[width=\textwidth]{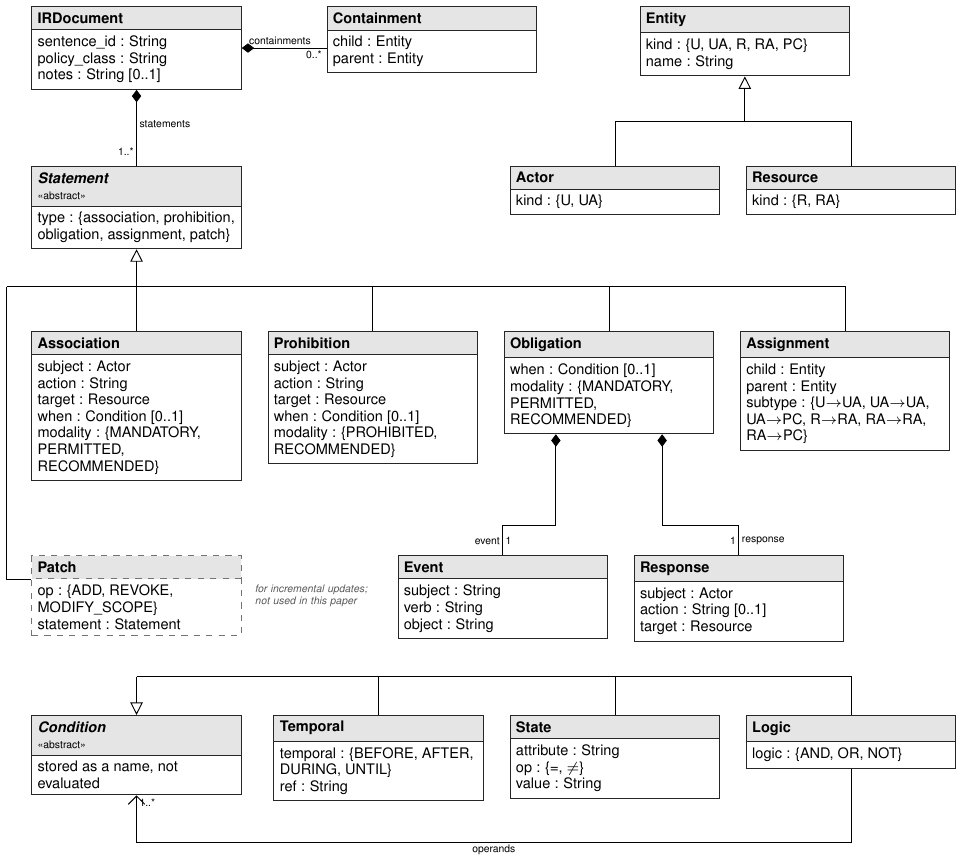}
\caption{The \sys{} metamodel (grammar v1.2) as a UML class diagram, drawn from the JSON Schema that restricts the LLM's output. Patch statements serve incremental updates and are not used in this paper.}
\label{fig:metamodel}
\end{figure}

\paragraph{The form} \Cref{fig:metamodel} shows the metamodel, which fixes what an IR document may contain. Each sentence becomes one IR document. It records the sentence identifier and policy class, one or more \emph{statements}, which are the rules of the sentence, and, optionally, \emph{containments}: simple child--parent pairs that the sentence implies, such as ``the system contains the patient list''.

A statement is one of four kinds. An \emph{association} says that a subject (a user or user attribute) may perform an action on a target (a resource or resource attribute). A \emph{prohibition} has the same parts but says that the subject may not perform the action. An \emph{obligation} says that when an event occurs, the system performs a response. An \emph{assignment} places a child under a parent, with one of the six kinds of \cref{def:assign}.

Subjects and targets have separate types in the grammar, so the LLM cannot write a rule whose subject is a resource. This error is prevented, not found later. In the main run, all 1{,}036 associations and 50 prohibitions had a user attribute as subject and a resource attribute as target, as \cref{def:assoc,def:prohib} require.

Associations, prohibitions and obligations also record how strong they are (``must'', ``may'' or ``should''), and they can carry a condition. Only sensible combinations of strength are allowed, so a prohibition can be ``must not'' or ``should not'', but never ``may''. A condition is a point in time (\texttt{AFTER registration\_closed}), a state (\texttt{status = in\_transit}), or a combination of these with AND, OR and NOT. The IR only names the condition; it does not check whether it holds. This way, ``students can register'' and ``students cannot register after registration closes'' can both be stored without looking like a conflict. We added each of these three features (separate types, rule strength and named conditions) because at least one of 20 sample sentences, the design set, could not be represented correctly without it.

\paragraph{How the LLM fills it in} For each sentence, the LLM receives the sentence, its policy class, the two sentences before it, and six examples: four from the same dataset and two hand-written ones with conditions. The sentence being extracted is never one of its own examples. A grammar, the JSON Schema of the metamodel, restricts the LLM's output while it is generated, through the structured-output feature of vLLM 0.31.0 with its default backend setting \citep{vllm2026structured}. Every completed answer is therefore an IR document that fits the form. An answer can only fail if it is cut off at the length limit; with a limit of 4{,}000 output tokens this never happened, as the longest answer used about 1{,}500.

\subsection{Step 2: the compiler builds the NGAC graph}\label{sec:t2}

The compiler applies fixed rules, listed in \cref{tab:transformation}, and uses no LLM. Take the association ``doctors may read patient records''. The compiler makes a node for doctor and a node for patient record, unless they already exist, and assigns both to the policy class. It then draws an association edge from doctor to patient record with the action read. The rule's condition and strength are stored on this edge. An obligation becomes a row in an obligation table. Because the rules are fixed, the same IR always gives the same graph, and every node, edge and obligation lists the sentences that produced it.

As the last part of Step~2, the compiler runs four structural checks, V1--V4 in \cref{tab:transformation}, which make sure the graph is well formed. V1 checks that the assignment hierarchy has no cycle. V2 checks that every association and prohibition goes from a user or user attribute to a resource or resource attribute; valid IR cannot break this, so V2 is only a safety net. V3 looks for an association and a prohibition on the same pair with a shared action: with the same condition this is a conflict and an error, and with different conditions it is only a warning. V4 reports a statement that is repeated within one sentence.

\input{tables/transformation_rules}

\subsection{Step 3: export to the Policy Machine}\label{sec:t3}

The export writes the graph in Policy Machine Language, the input format of NIST's Policy Machine. Each policy class, attribute, user and resource becomes a \texttt{create} statement, and parents are written before their children; a node without a parent is placed under a policy class named \texttt{Unattached}. Each association becomes an \texttt{associate} statement and each prohibition a \texttt{deny} statement. Obligations are kept in a separate table, because our profile of the language has no statement for events. The language cannot express conditions or the strength of a rule, so both are kept as comments. Because the Policy Machine cannot check a condition, we also test a strict export in \cref{sec:rq1}, which leaves conditional permissions out, so they are denied. Rules of one user attribute on the same resource are merged into one statement, because the Policy Machine keeps only one association per pair. At this point the file is only a draft. A person first looks at the rules that the back-translation check flags (\cref{sec:s6}) and then either accepts the draft or corrects it before it is enforced.

\subsection{The back-translation check}\label{sec:s6}

The structural checks V1--V4 of Step~2 can only show that the graph is well formed. A rule that names the wrong resource still passes them, as long as that resource exists in the graph. The back-translation check therefore looks at meaning. It runs on the IR and the graph as soon as Step~2 is done, so the exported file can be made at the same time. It has four steps: rule checks, Write, Judge and Flag. We call Write and Judge together \emph{the LLM step}, because the judge is an LLM.

\paragraph{Rule checks} Five checks that use no LLM look at each IR document and catch cases the LLM step cannot see. C1 flags a circular obligation, whose event repeats its own response (for example ``when the system notifies the billing system, the system notifies the billing system''). C2 flags a permission or prohibition whose action is a state verb such as \emph{has} or \emph{is}. C3 flags an empty source sentence. C4 flags a sentence that states a permission or duty (\emph{can}, \emph{may}, \emph{must}, \emph{shall}, \emph{will}, \emph{should}) when the IR records only containments. C5 flags a sentence in which someone ``is presented with'', ``is provided with'', ``is shown'' or ``is given'' something, which states a permission to view it, when the IR records only containments.

\paragraph{Write} Each rule that grants, denies or requires an action is written as one plain English sentence from a fixed template, together with its condition when it has one.

\paragraph{Judge} A separate LLM call, the \emph{judge}, receives the written rule, its source sentence, found through the rule's sentence identifiers, and the two sentences before it. The judge may use the earlier sentences only to resolve words such as ``he'' or ``the user'' and the actor of a passive sentence; a rule that only an earlier sentence states counts as a mismatch. It answers whether the sentence states this rule with the same resource, or with a more general, more specific or different one, and gives a one-line reason. Because a sentence often states several rules, the judge is told that the rule only needs to be one of them.

\paragraph{Flag} A rule check that fires, or a mismatch from the judge, flags the rule or sentence. A person then looks at each flag together with its sentence, the reason and the matching statement in the exported Policy Machine file, and decides what to do. If the rule is right, it passes to enforcement unchanged. If it is wrong, the person corrects or removes it in the IR of that sentence, so only the flagged sentences change, and Steps~2 and~3 of the pipeline run again, which takes seconds and needs no LLM. Once the person has cleared every flag, the exported file is final, and only then is it loaded into the Policy Machine and enforced.

Neither the rule checks nor the judge changes the output, and neither sees the ground truth or the rest of the graph.

\section{Evaluation setup}\label{sec:setup}

We ask three research questions.
\textbf{RQ1 (valid and enforceable output):} does constrained decoding make every extraction compilable, at what cost in accuracy, and can the compiled policy be enforced?
\textbf{RQ2 (extraction quality):} how faithful is \sys{}'s output to its source sentences compared with published and rebuilt baselines, and why does exact match understate this?
\textbf{RQ3 (finding errors of meaning):} how many errors of meaning does the back-translation check detect, how many sentences does it send to review, and how precisely do its flags point to the error? RQ1 tests the first contribution, RQ3 the second, and RQ2 the third and fourth: it measures faithfulness for all methods and compares \sys{}, which runs at zero API cost, with Text2NGAC, which uses paid commercial LLMs.

\paragraph{Datasets} We use the same four datasets as Text2NGAC \citep{habib2026text2ngac}, all taken from public requirements documents: the iTrust health-record system \citep{meneely2012itrust} in the versions used by Text2Policy \citep{xiao2012automated} (T2P, 234 policy sentences) and by ACRE \citep{slankas2014relation} (351), the IBM course-registration system \citep{ibm2010course} (IBM, 75) and the CyberChair conference system (CC, 54), 714 policy sentences in total. We use the NGAC ground truth released with Text2NGAC, which covers all four relation types. The grammar was designed with the 20 sentences of the design set (\cref{sec:metamodel}), which are also in the scored set. To check that they do not favor \sys{}, we scored without them, and exact match stays the same.

\paragraph{Models and hardware} The main extractor is Qwen3.8-27B (the official version with 8-bit weights, FP8). All results use this model unless stated otherwise, and every rebuilt baseline uses it too. To check that the results do not depend on this one model, we repeat the comparison with the closest baseline, direct extraction, on gpt-oss-20b, a model from another family: we run both \sys{} and direct extraction with gpt-oss-20b and compare them in exact match and in faithfulness (\cref{sec:rq2}). The judge in the back-translation check is Qwen3.8-27B, in a call separate from extraction; we repeat the check with gpt-oss-20b as the judge. All runs used open-weight models through vLLM, with temperature 0 and a fixed seed, on one local server with two NVIDIA RTX 4090 GPUs (24\,GB each); no commercial API was used. A full extraction run over all 714 sentences takes about 20 minutes on this server, including loading the model (about 1.7\,s per sentence; requests run in parallel), and the back-translation check of all 1{,}148 extracted rules takes about 3 minutes. At the GPUs' rated power of 450\,W each, this is at most about 0.35\,kWh of GPU energy for the whole pipeline. For comparison, Text2NGAC needs 0.55--0.64 minutes per sentence and \$11--47 in API fees per 1{,}000 sentences \citep{habib2026text2ngac}. Even with temperature 0 and a fixed seed, results can change slightly from run to run, because the GPU processes many requests in parallel. To measure this, we ran the extraction three times with the same settings on the 234 sentences of the T2P dataset: the number of sentences that match the ground truth exactly differed by at most 2 between the runs.

\paragraph{Baselines} (a) \emph{No grammar}: the same model, prompt and context in plain JSON mode, isolating the effect of constrained decoding. (b) \emph{Direct extraction}: a re-implementation of the single-model extraction step of our earlier system Text2NGAC \citep{habib2026text2ngac}, which writes NGAC tuples directly, without the IR. It needs the relation type of each sentence, which we give in two ways: predicted by one LLM call (realistic), or taken from the ground truth (an upper bound). (c) Two published methods, the rule-based pipeline of \citet{abdelgawad2023synthesizing} and Code4Policy \citep{lawal2024translating} (\cref{sec:rq2}). (d) The released outputs of the full Text2NGAC pipeline, which uses several commercial LLMs; we compare with it on faithfulness only, which does not use the ground truth. (e) For RQ3, two checks from Text2NGAC: its structural verifier, which only tests whether the graph is well formed, and its parent-abstraction detector, which flags a rule when the sentence names a more specific resource than the rule does.

\paragraph{Main metric: faithfulness} A sentence's output is faithful if every extracted rule says what the sentence says, using the preceding sentences only to resolve words such as ``he'' or ``the user''. Each output was marked against its sentence, without the ground truth, as \emph{correct}; \emph{minor problem} (the main permission or duty is right, but a name is too general or too specific, a small detail is missing, or the actor of a passive sentence is guessed); or \emph{error} (wrong actor, action, object or trigger; permission and prohibition swapped; a stated permission or duty recorded only as a containment; the main point missing; or a rule with no basis in the text). To mark fairly, we marked the outputs of all systems blind: each output was shown without the name of the system that produced it and in shuffled order, and when two systems gave the same output for a sentence, it was marked once, so both received the same mark. One reading was applied to all systems alike: a sentence in which a named person is presented with or provided with something, such as ``The LHCP is presented with a list of other appointments that conflict with the new one'', gives that person permission to view it, so an output that records only a containment (an assignment that places the list under the system) is an error. When only the system shows something and no person is named, as in ``System presents an automatically-calculated value of BMI'', a containment alone is correct. The authors did the marking, assisted by an LLM, following a written guide that we release with the marks. A sentence counts as faithful if its output is marked correct or has only a minor problem. We also report a stricter measure: the share of sentences whose output is correct with no problem at all.

\paragraph{Why not exact match} Exact match counts a sentence as correct only if the compiled tuples equal the ground-truth tuples after the same name normalization as in Text2NGAC \citep{habib2026text2ngac}. A policy sentence can often be written correctly in more than one way, and annotators make convention choices (\texttt{appointment\_detail} or \texttt{appointment\_details}; whether ``the system displays X'' is an assignment or an obligation) that exact match counts as errors. Two careful annotators would also differ: on these datasets, an independent second annotator did not always give the same rule type as the first ($\kappa = 0.84$) \citep{habib2026text2ngac}, and exact match on the full rules also needs the same name for every actor, action and resource. We report exact match only to compare with earlier work, which was scored this way.

\paragraph{Inserted errors for RQ3} To test whether the back-translation check finds wrong rules, we also insert errors on purpose. A \emph{mutant} is a copy of a correct graph with one such error. We reuse the 120 mutants of Text2NGAC \citep{habib2026text2ngac}, in which one permission names a broader resource than its sentence does: the resource is replaced by its parent in the hierarchy, for example ``patient record'' instead of ``patient MID''. The rule keeps its link to the source sentence. To see how often the check flags a correct rule, we also run it on the 610 association edges that we leave unchanged and that are linked to a sentence.

\section{Results}\label{sec:results}

\subsection{RQ1: valid and enforceable output}\label{sec:rq1}

\Cref{tab:rq1} compares \sys{} with and without the grammar. Both runs use the same model, prompt, examples and context; the only difference is that without the grammar the LLM writes plain JSON. Without the grammar, 32 of the 714 outputs (4.5\%) cannot be compiled. In 26 of them a part is missing or of the wrong kind, for example an obligation without a response, or a resource used as the subject of a rule. The other 6 contain fields that the IR does not have. With the grammar, all 714 outputs fit the IR and can be compiled, and the compiler's type check (V2) never found an error.

\paragraph{Loading into the NIST Policy Machine} To check that the compiled policies can be enforced, we loaded the Policy Machine Language output for all four datasets into NIST's reference implementation, \texttt{policy-machine-core} 3.3.0 \citep{ferraiolo2015pm}, as was done for Text2NGAC \citep{habib2026text2ngac}. The Policy Machine accepted every statement. Of the 1{,}880 edges in the four compiled graphs, the export left out only two containments that NGAC does not allow, such as a user attribute placed inside a resource attribute. The Policy Machine needs every attribute to belong to a policy class. Most attributes get one through a rule of their sentence, but an attribute that appears only in a containment gets none. For example, ``System presents an automatically-calculated value of BMI'' produces only a containment that links the BMI value to the system, and no rule that gives anyone access to it, so the BMI value belongs to no policy class. This happened for 54 attributes in T2P, 4 in CC and 73 in ACRE; the export places all of them in one extra policy class, as Text2NGAC did. In addition, the compiled policies name roles, such as HCP (health care personnel), but no actual users, while the Policy Machine decides requests for users. We therefore added one test user to each user attribute, for example a test user named \texttt{u\_demo\_0} who is an HCP, so that we could send access requests on that user's behalf. We then sent 390 access requests (T2P 98, IBM 86, CC 84, ACRE 122): permissions granted directly or through inheritance, denials by prohibitions, and requests with no permission. For each request, we worked out the expected answer separately from the NGAC rules on the compiled graph. The Policy Machine gave the expected answer to all 390. We also registered one obligation per dataset and sent its event, and all four responses were carried out. This shows that the Policy Machine enforces the compiled graph as written; whether the graph itself is correct is measured by faithfulness (\cref{sec:rq2}).

\paragraph{Conditions and rule strength in enforcement} The Policy Machine cannot check conditions, so the export above grants a conditional permission at all times. This gives more access than the policy states for 98 of the 715 permissions (13.7\%). We therefore also made a strict export, which leaves conditional permissions out of the Policy Machine and lists them, with their condition and sentence, for the application to check, for example whether registration has closed. With the strict export, the Policy Machine again gave the expected answer to all 390 requests: 23 requests that the first export grants are now denied, no denial changes, and the obligations are still carried out. Conditional prohibitions are kept and always apply, which can only deny more access, never grant more. ``Must'' and ``may'' are both exported as permissions. The Policy Machine does not check that a duty is carried out, but the IR keeps the strength of every rule, so these rules can be passed to an audit. For deployment we recommend the strict export.

\begin{table}[!tbp]
\centering\small
\caption{RQ1. Validity and accuracy with and without constrained decoding (same model, prompt and context). For plain JSON, outputs that cannot be compiled count as failures in every metric.}
\label{tab:rq1}
\begin{tabular}{llccccc}
\toprule
Metric & Decoding & T2P & IBM & CC & ACRE & All \\
\midrule
\multirow{2}{*}{Compilable (\%)} & plain JSON & 98.3 & 96.0 & 92.6 & 94.0 & 95.5 \\
 & grammar (\sys) & 100 & 100 & 100 & 100 & 100 \\
\midrule
\multirow{2}{*}{Exact match (\%)} & plain JSON & 53.4 & 64.0 & 31.5 & 35.6 & 44.1 \\
 & grammar (\sys) & 53.0 & 64.0 & 29.6 & 37.0 & 44.5 \\
\bottomrule
\end{tabular}
\end{table}

\paragraph{Accuracy} The grammar does not change accuracy: exact match over all 714 sentences is 44.1\% without it and 44.5\% with it. What the grammar adds is that every output is valid. The compiler needs this to build the graph and keep the source sentence of each rule; without the grammar, about one output in twenty cannot be compiled and is lost.

\subsection{RQ2: extraction quality}\label{sec:rq2}

\paragraph{Faithfulness} \Cref{tab:faithful} shows how often the output of each system says what its sentence says. \sys{} is faithful on 95.7\% of sentences, from 93.4\% on ACRE to 100\% on CC, while the other systems in the upper part of the table reach 83.9--90.3\%. \sys{} also makes the fewest errors, only 31, whereas the others make 69 to 115. It also wins on the stricter measure: 77.0\% of its outputs are correct with no problem, whereas the others reach at most 73.0\%.

To check that this lead is not due to chance, we compared \sys{} with each baseline sentence by sentence and counted the sentences on which only one of the two is faithful. Only \sys{} is faithful on 46 sentences, whereas only Text2NGAC is faithful on 7. The gap is similar or larger for the other baselines: \sys{} has 52 such sentences while direct extraction with the ground-truth type has 14, \sys{} has 62 while direct extraction with the predicted type has 18, and \sys{} has 108 while Code4Policy has 24. McNemar's test confirms that none of these gaps is due to chance ($p < 10^{-5}$).

Most of \sys{}'s 31 errors are of one kind: in 18 of them, a permission or duty that the sentence states is recorded only as a containment. The others give an obligation a wrong trigger (7), drop a condition (4), name a wrong object (1), or come from an empty sentence (1).

\begin{table}[!tbp]
\centering\small
\caption{RQ2. Faithfulness on all 714 sentences. Upper part: Qwen3.8-27B, and the released outputs of Text2NGAC. Lower part: gpt-oss-20b.}
\label{tab:faithful}
\setlength{\tabcolsep}{4pt}\begin{tabular}{llccccc}
\toprule
 & System & T2P & IBM & CC & ACRE & All \\
\midrule
 & Sentences & 234 & 75 & 54 & 351 & 714 \\
\midrule
\multirow{5}{*}{Faithful (\%)} & \sys{} & \textbf{97.4} & 97.3 & \textbf{100} & \textbf{93.4} & \textbf{95.7} \\
 & Direct, ground-truth type & 96.2 & 90.7 & 79.6 & 88.0 & 90.3 \\
 & Text2NGAC \citep{habib2026text2ngac} & 96.6 & \textbf{100} & 83.3 & 84.9 & 90.2 \\
 & Direct, predicted type & \textbf{97.4} & 84.0 & 85.2 & 86.0 & 89.5 \\
 & Code4Policy \citep{lawal2024translating} & 91.5 & 82.7 & 81.5 & 79.5 & 83.9 \\
\midrule
\multirow{5}{*}{\shortstack[l]{Correct, no\\problem (\%)}} & \sys{} & 88.0 & 73.3 & \textbf{64.8} & \textbf{72.4} & \textbf{77.0} \\
 & Direct, ground-truth type & 87.2 & 65.3 & 50.0 & 65.2 & 71.3 \\
 & Text2NGAC \citep{habib2026text2ngac} & \textbf{88.9} & \textbf{76.0} & 53.7 & 64.7 & 73.0 \\
 & Direct, predicted type & 87.2 & 60.0 & 51.9 & 62.7 & 69.6 \\
 & Code4Policy \citep{lawal2024translating} & 62.0 & 38.7 & 37.0 & 49.9 & 51.7 \\
\midrule
\multicolumn{7}{@{}l}{\emph{With gpt-oss-20b}} \\
\multirow{2}{*}{Faithful (\%)} & \sys{} & \textbf{95.3} & \textbf{94.7} & \textbf{88.9} & \textbf{89.2} & \textbf{91.7} \\
 & Direct, predicted type & 88.5 & 74.7 & 74.1 & 76.6 & 80.1 \\
\cmidrule(l){2-7}
\multirow{2}{*}{\shortstack[l]{Correct, no\\problem (\%)}} & \sys{} & \textbf{84.6} & \textbf{70.7} & \textbf{51.9} & \textbf{65.8} & \textbf{71.4} \\
 & Direct, predicted type & 74.4 & 53.3 & 37.0 & 57.8 & 61.2 \\
\bottomrule
\end{tabular}
\end{table}

\paragraph{Against published methods} \Cref{tab:published} compares \sys{} with two published extractors on the same four datasets. The rule-based pipeline of \citet{abdelgawad2023synthesizing} predicts only the type of rule that a sentence contains, so it can be scored on rule type but not on exact match or faithfulness. Code4Policy \citep{lawal2024translating} asks an LLM to fill in a fixed, code-like template of permissions and denials; we ran it with its published prompt on the same model as \sys{}. Rule type counts a sentence as correct if the predicted rules include its annotated type. For Code4Policy's exact match we use its own, more lenient scoring, which groups verbs into read, write, create, delete and execute; with \sys{}'s scoring it would reach only 1.5\%.

\sys{} has the highest exact match on every dataset and beats the rule-based pipeline on rule type everywhere. Code4Policy matches \sys{}'s rule type only on IBM, where 85\% of sentences are associations. Its template cannot express obligations or containments, so it gets every sentence that needs one wrong, and it is less faithful than \sys{} on every dataset (\cref{tab:faithful}).

\begin{table}[!tbp]
\centering\small
\caption{RQ2. \sys{} and two published methods on all four datasets. Code4Policy's exact match uses its own, more lenient scoring.}
\label{tab:published}
\begin{tabular}{llcccc}
\toprule
Metric & System & T2P & IBM & CC & ACRE \\
\midrule
\multirow{3}{*}{Rule type (\%)} & Rule-based \citep{abdelgawad2023synthesizing} & 73.9 & 46.7 & 33.3 & 59.3 \\
 & Code4Policy \citep{lawal2024translating} & 76.5 & \textbf{90.7} & 53.7 & 68.1 \\
 & \sys{} & \textbf{97.4} & \textbf{90.7} & \textbf{72.2} & \textbf{90.9} \\
\midrule
\multirow{2}{*}{Exact match (\%)} & Code4Policy \citep{lawal2024translating} & 17.9 & 18.7 & 3.7 & 12.5 \\
 & \sys{} & \textbf{53.0} & \textbf{64.0} & \textbf{29.6} & \textbf{37.0} \\
\bottomrule
\end{tabular}
\end{table}

\paragraph{Against direct extraction} Direct extraction is the closest baseline: the same model writes NGAC tuples directly, without the IR. In exact match (\cref{tab:rq2}), \sys{} is ahead of it on T2P, IBM and ACRE, by 3.0 to 14.7 points with a realistic predicted type. It is behind only on CC, a small dataset (54 sentences) whose annotation follows conventions that the direct prompt's examples copy; in faithfulness this gap disappears, as \sys{} is faithful on all 54 CC sentences whereas the direct extractor is faithful on 85.2\%. The same holds with gpt-oss-20b: \sys{} leads in exact match on T2P, IBM and ACRE, and in faithfulness on all four datasets, with 91.7\% whereas direct extraction reaches 80.1\% (lower part of \cref{tab:faithful}; McNemar's test, $p < 10^{-5}$). The gain comes from the IR, not from the grammar: without the grammar, the IR run still reaches 44.1\% exact match over all sentences, whereas the direct extractor reaches 40.9\% (\cref{tab:rq1}).

\begin{table}[!tbp]
\centering\small
\caption{RQ2. Exact match (\%) of \sys{} and direct extraction, for two model families. Direct extraction is given the relation type of each sentence, either predicted by an LLM call or taken from the ground truth (an upper bound).}
\label{tab:rq2}
\begin{tabular}{llcccc}
\toprule
Model & System & T2P & IBM & CC & ACRE \\
\midrule
\multirow{3}{*}{Qwen3.8-27B} & \sys{} & \textbf{53.0} & \textbf{64.0} & 29.6 & \textbf{37.0} \\
 & Direct, predicted type & 50.0 & 49.3 & \textbf{63.0} & 29.6 \\
 & Direct, ground-truth type & 53.0 & 64.0 & 74.1 & 30.5 \\
\midrule
\multirow{2}{*}{gpt-oss-20b} & \sys{} & \textbf{42.7} & \textbf{52.0} & 18.5 & \textbf{33.3} \\
 & Direct, predicted type & 32.1 & 42.7 & \textbf{35.2} & 25.4 \\
\bottomrule
\end{tabular}
\end{table}

\paragraph{Exact match understates quality} Exact match and faithfulness rank the systems differently. Given the ground-truth type, the direct extractor matches the annotation exactly on 44.7\% of sentences, as often as \sys{} (44.5\%), but it is faithful on only 90.3\%, whereas \sys{} is faithful on 95.7\%. On CC it matches the annotation on 74.1\% of sentences and \sys{} on 29.6\%, yet \sys{} is faithful on all 54 and the direct extractor on 79.6\%. Exact match rewards following the annotator's conventions, not reading the sentence correctly. Even two careful annotators can write the same rule differently (\cref{sec:setup}), so a mismatch with one annotation does not mean that a rule is wrong. For this reason, faithfulness is our main measure of quality: it checks whether each extracted rule says what its sentence says.

\subsection{RQ3: finding errors of meaning}\label{sec:rq3}

We first test the LLM step on the inserted errors described in \cref{sec:setup}: 120 rules that name a broader resource than their sentence, for example ``patient record'' where the sentence says ``patient MID''. These are the errors that Text2NGAC's own checks were built for. Text2NGAC's structural verifier finds only 11 of them (9.2\%), because a rule with a broader resource is still well formed. Its second, special check finds 95.0\%, but it was written for this one kind of error: it flags a rule only when the sentence names a more specific resource that sits below the rule's resource in the hierarchy. The LLM step has no special logic for this error; it simply asks whether the sentence states the rule. It finds 109 of the 120 errors (90.8\%), close to the special check, and it flags only 2.3\% of the 610 correct association rules, as few as the special check (2.3--5.3\%).

\paragraph{More kinds of error and all kinds of rule} A broader resource is only one possible error, and associations are only one kind of rule. We therefore extended the test in two ways (\cref{tab:rq3b}). First, we inserted five more kinds of error into the same graphs: a wrong resource, a wrong actor, a wrong action, a permission turned into a prohibition or the reverse, and a wrong response in an obligation. Each wrong name shares no word with the original and does not appear in the sentence. Second, we also checked prohibitions and obligations, not only associations.

\begin{table}[!tbp]
\centering\small
\caption{RQ3. Share of inserted errors found, per kind of error, and share of correct rules flagged, per kind of rule, on the four Text2NGAC graphs \citep{habib2026text2ngac}. The judge sees the two preceding sentences as context.}
\label{tab:rq3b}
\begin{tabular}{lrc}
\toprule
Kind of error or rule & Cases & Found or flagged (\%) \\
\midrule
\multicolumn{3}{l}{\emph{Inserted errors found}} \\
Broader resource & 120 & 90.8 \\
Wrong resource & 120 & 98.3 \\
Wrong actor & 120 & 95.0 \\
Wrong action & 120 & 86.7 \\
Permission and prohibition swapped & 120 & 100 \\
Wrong obligation response & 50 & 100 \\
\midrule
\multicolumn{3}{l}{\emph{Correct rules flagged}} \\
Associations & 610 & 2.3 \\
Prohibitions & 23 & 8.7 \\
Obligations & 58 & 25.9 \\
\bottomrule
\end{tabular}
\end{table}

\Cref{tab:rq3b} shows that the LLM step finds most errors of every kind: between 86.7\% and 100\%, and 94.6\% of all 650. It also rarely flags a correct rule: 2.3\% of correct associations and 8.7\% of correct prohibitions (two of 23). Two results look weaker, but partly for reasons outside the check. First, several of the 11 broader-resource errors that it misses do not seem to be real errors. For example, one recorded change replaced \texttt{notification} with \texttt{acceptance} in a sentence about ``notifications of acceptance and rejection'', so the sentence still supports the changed rule. Second, it flags 15 of the 58 correct obligations (25.9\%), but many of these flags seem to point to real mistakes in the Text2NGAC graph \citep{habib2026text2ngac} rather than false alarms. For example, one obligation sends a message to the system where the sentence sends it to the authors, and another turns a request to a person (``reviewers are asked to inform the maintainer'') into an automatic action.

\paragraph{Real extraction errors} The inserted errors are made up, so we also ran the back-translation check on the real output of \sys{}: all 714 sentences and their 1{,}148 rules. The check sends 75 sentences (10.5\%) to a person for review. These 75 include 26 of the 31 sentences marked as errors in \cref{tab:faithful} (84\%) and 24 of the 133 with a minor problem; the other 25 are correct. The result is similar on both halves of the data: 12 of 14 errors are found in one half and 14 of 17 in the other.

The LLM step and the rule checks catch different errors. The most common error, a permission or duty recorded only as a containment, gives the LLM step no rule to judge, and this is exactly the case the rule checks are written for: C5 alone flags 11 sentences, 10 of which are errors. The 5 errors that are not found are all well-formed rules that stay close to the wording of the sentence: 3 lose a condition, one names a wrong object, and one is a list item (``a descriptive name for the instructions'') whose permission is stated only in an earlier sentence. The reviewed flags are part of the artifact.

\paragraph{A judge from another model family} The judge above is the same model as the extractor, so the two may share blind spots. We therefore repeated the LLM step with gpt-oss-20b as the judge, with the same prompt and context. Together with the rule checks, it sends 73 of the 714 sentences (10.2\%) to review and finds 24 of the 31 errors, whereas Qwen3.8-27B sends 75 sentences and finds 26 errors. Used together, the two judges miss only 4 errors. On the inserted errors of \cref{tab:rq3b}, gpt-oss-20b flags fewer correct rules (3.8\% of the 691), while Qwen3.8-27B flags 4.5\%; Qwen3.8-27B finds more of the inserted errors (94.6\%), whereas gpt-oss-20b finds 85.1\%. So the check does not depend on one judge model, and Qwen3.8-27B is the stronger of the two.

\paragraph{Localization} A flag is only useful if a reviewer can quickly see what is wrong. We measured this on the 120 broader-resource errors and the 610 correct associations. Each flag points to the sentences its rule came from: 2.5 sentences on average, and exactly one sentence for 56\% of flags. Checking every flag, right or wrong, means rereading 15.7\% of the sentences, whereas the structural checks point to no sentence at all. The judge's one-line reason also helps: for 81\% of the errors found, it names both the wrong resource and the one in the sentence, so the reviewer can see the fix at once. This share is lowest on T2P and ACRE (61\% and 64\%), but even there most reasons name at least one of the two resources; only 4 of the 109 reasons name neither.

\section{Discussion}\label{sec:discussion}

\paragraph{What the grammar gives} The grammar does not make the model more accurate; it makes the steps after the model reliable. Every complete IR document can be compiled, the compiler gives the same graph for the same input, and every rule keeps the sentence it came from. An auditor needs these properties, and they hold for every output, not only on average.

\paragraph{Where the remaining errors are} The remaining errors (\cref{sec:rq2}) are errors of reading, not of format. Most of them (18 of 31) record a stated permission or duty only as a containment, and the back-translation check, mainly through C4 and C5, finds 17 of these. The errors it does not find lose a condition or name a wrong object while staying close to the wording of the sentence. Finding them needs a check that compares conditions and names with the sentence.

\paragraph{Threats to validity} \emph{Internal.} Faithfulness depends on how each output was marked. All outputs were marked blind, without system names and in shuffled order, so a difference between systems cannot come from knowing which system produced an output. The marks still come from a single marking pass with one written guide, and the line between a minor problem and an error depends on that guide; we release the marks and the guide so that others can repeat the marking. The judge is the same model as the main extractor (Qwen3.8-27B). It runs in a separate call that sees neither the IR nor the extraction prompt, but the two may still share blind spots. With a judge from another model family (gpt-oss-20b), the check finds 24 instead of 26 of the 31 errors (\cref{sec:rq3}). \emph{Construct.} Most errors in RQ3 are inserted on purpose. We cover six kinds of error and also test the check on real output, but five of the six kinds are our own, and the unchanged rules of the Text2NGAC graphs \citep{habib2026text2ngac} are not all correct, so some flags counted as false may in fact be right. \emph{External.} The four datasets come from three domains and are the standard public benchmarks; real policy documents may be longer and less regular. \emph{Reproducibility.} All runs use open-weight models at temperature 0 on one server with two GPUs, and repeated runs differ by at most 2 of 234 sentences. The grammar, compiler, checks, prompts, outputs and scoring scripts will be released publicly on publication.

\paragraph{Limitations} The grammar cannot express quantities, delegation or nested obligations. The Policy Machine does not check conditions; the strict export (\cref{sec:rq1}) leaves conditional permissions to the application, and duties stated with ``must'' are enforced only as permissions. Obligations are kept in a separate table; the export does not load them into the Policy Machine, and our test registered only one obligation per dataset. The back-translation check is weakest on obligations, because an event and its response are harder to write as one clear sentence.

\section{Conclusion}\label{sec:conclusion}

We turned natural-language-to-NGAC translation into a model-driven pipeline. The language model only fills in a fixed template (the IR), and a grammar keeps its output within this template. A compiler with fixed rules builds the NGAC graph and keeps a link to each source sentence, and a back-translation check compares each rule with that sentence. The grammar removes broken output without changing accuracy. The extracted rules are faithful to their sentence for 95.7\% of sentences, whereas Code4Policy, our earlier system and a direct extractor reach at most 90.3\%, even where exact match with the annotation is low. The back-translation check finds 94.6\% of the inserted errors of meaning that structural tests miss, finds 26 of the 31 real errors, and points each flag to the sentence to reread. All of this runs on open-weight models on one local server, with zero API cost, where our earlier system needed an ensemble of commercial LLMs. Future work will improve the check on obligations, test it on independently written policies, and use the IR for incremental policy updates. We also observed that some reference labels name a resource more generally than the sentence does: for example, the sentence ``patients can view their upcoming appointments'' is labeled with the resource \texttt{appointment}, which would also cover past appointments. In our datasets such cases are rare (18 of 714 sentences) and do not affect the results, but in a deployed system a label like this would grant more access than the policy states. Future work will define a naming rule that keeps every word that narrows the resource, and apply it to both extraction and reference labels.

\section*{Data availability}
The grammar, compiler, checks, prompts, the outputs of all systems, their faithfulness marks with the marking guide, the reviewed back-translation flags, the Policy Machine test, and the scoring scripts will be made publicly available upon publication of this article.

\section*{Declaration of generative AI and AI-assisted technologies in the writing process}
During the preparation of this work the authors used Claude (Anthropic) to assist with programming the experiment scripts and with drafting and editing the text. After using this tool, the authors reviewed and edited the content as needed and take full responsibility for the content of the publication. The language models used as components of the method (Qwen and gpt-oss) are described in \cref{sec:setup}.

\bibliographystyle{elsarticle-num-names}
\bibliography{references}
\end{document}

%% file: tables/transformation_rules.tex
\begin{table}[!tbp]
\centering\footnotesize
\caption{Rules of $T_2$, from IR constructs to NGAC graph elements (R1--R8), and the four checks that run after it (V1--V4). Every node, edge and obligation keeps the \texttt{sentence\_id} of the IR document that produced it. $PC$ is the policy class of the document.}
\label{tab:transformation}
\renewcommand{\arraystretch}{1.05}
\begin{tabularx}{\textwidth}{@{}l >{\raggedright\arraybackslash}p{2.9cm} >{\raggedright\arraybackslash}X >{\raggedright\arraybackslash}p{2.6cm}@{}}
\toprule
\textbf{Rule} & \textbf{IR construct} & \textbf{NGAC elements produced} & \textbf{Condition or report} \\
\midrule
R1 & \texttt{IRDocument}(\textit{sid}, \textit{pc}) & node $PC{:}pc$; \textit{sid} is recorded as the source of everything below & one document per sentence \\
R2 & \texttt{Containment}(\textit{child}, \textit{parent}) & nodes for both entities; assignment $child \to parent$, typed by their kinds & possessives, ``X of Y'' \\
R3 & \texttt{Association}(\textit{s}, \textit{a}, \textit{t}, \textit{when}, \textit{m}) & nodes $s$, $t$; assignments $s\to PC$, $t \to PC$; association edge $(s,t)$ whose operation set gains $a$; condition and modality $m$ stored on the edge & $m$ = RECOMMENDED: warning \\
R4 & \texttt{Prohibition}(\textit{s}, \textit{a}, \textit{t}, \textit{when}, \textit{m}) & as R3, but a prohibition edge & as R3 \\
R5 & \texttt{Obligation}(\textit{event}, \textit{response}, \textit{when}, \textit{m}) & nodes for the response's subject and target, with their assignments to $PC$; an obligation row (event $\Rightarrow$ response), condition, $m$ & $m \neq$ MANDATORY: warning \\
R6 & \texttt{Assignment}(\textit{child}, \textit{parent}, \textit{subtype}) & nodes; assignment $child\to parent$ of the given subtype & one of the six kinds (enforced by the grammar) \\
R7 & \texttt{Condition} (time, state or logic) & stored as text on the edge, e.g.\ \texttt{AFTER(ref)}, \texttt{attr=value}, \texttt{AND(\ldots)}; exported as a comment & lets otherwise conflicting edges coexist (V3) \\
R8 & \texttt{Patch}(\textit{op}, \textit{statement}) & not compiled in batch mode & error (incremental updates, future work) \\
\midrule
\multicolumn{4}{@{}l}{\textbf{Checks after the transformation} (reported with the sentence identifiers)} \\
V1 & assignment edges & the assignment relation must have no cycle & error, cycle listed \\
V2 & association and prohibition edges & source in $\{U,UA\}$, target in $\{R,RA\}$ & error (cannot occur with valid IR) \\
V3 & association and prohibition on the same pair with shared operations & same condition: conflict (error); different conditions: override (warning) & conflict detection \\
V4 & repeated statement in one sentence & duplicate statement & warning \\
\bottomrule
\end{tabularx}
\end{table}